\documentclass[a4paper,11pt]{article}
\pdfoutput=1 

\usepackage{jinstpub} 
\usepackage{graphicx}
\usepackage{subfigure}
\usepackage{verbatim}

\title{\boldmath The Automatic Calibration Unit in JUNO}

\author[a]{Jiaqi Hui,}
\author[b]{Hancheng Liu,}
\author[a,c]{Jianglai Liu,}
\author[a]{Yue Meng,}
\author[d,e,1]{Mengjiao Xiao,\note{Corresponding author, mengjiaoxiao@gmail.com}}
\author[c,a]{Donglian Xu,}
\author[b]{Lei Yang,}
\author[c]{Ziping Ye,}
\author[a]{Feiyang Zhang,}
\author[a,2]{Tao Zhang,\note{Corresponding author, tzhang@sjtu.edu.cn}}
\author[a]{Yuanyuan Zhang}

\affiliation[a]{School of Physics and Astronomy, Shanghai Jiao Tong University, 
MOE Key Laboratory for Particle Astrophysics and Cosmology, 
Shanghai Key Laboratory for Particle Physics and Cosmology, Shanghai 200240, China}
\affiliation[b]{School of Electrical Engineering and Intelligent, Dongguan University of Technology, Dongguan 523808, China}
\affiliation[c]{Tsung-Dao Lee Institute, Shanghai Jiao Tong
  University, Shanghai 200240, China} 
\affiliation[d]{Department of
  Physics, University of Maryland, College Park, MD 20742, USA}
\affiliation[e]{Center of High Energy Physics, Peking University,
  Beijing 100871, China}

\abstract{This paper describes the design and construction of the
  automatic calibration unit (ACU) for the JUNO experiment. The ACU is a
  fully automated mechanical system. It is capable of deploying multiple
  radioactive sources, an ultraviolet (UV) laser source, or an auxiliary sensor such as a temperature sensor,  one at a time, into the
  central detector of JUNO along the central axis. It is designed as a
  primary tool to precisely calibrate the energy scale of detector,
  aligning timing for the photosensors, and partially monitoring the
  position-dependent energy scale variations.}


\begin{document}
\maketitle
\flushbottom


\section{Introduction}
JUNO is an ultra-low background liquid scintillator (LS) detector under
construction in the Jiangmen city in south China~\cite{ref:JUNO_CDR}. At present, the civil construction is mostly completed, and the completion of the detector assembly is expected by the end of 2022. It is designed as a
general low background terrestrial and astrophysical neutrino
observatory, particularly to perform precision oscillation studies
using reactor neutrinos at a baseline of 53 km, in order to determine
the neutrino mass hierarchy (MH)~\cite{ref:baseline}. Its central detector (CD) is an
acrylic spherical vessel with an inner diameter of 35.4 m, viewed by
about 17,600 20-inch and 25,600 3-inch photomultiplier tubes (PMTs)
immersed in an ultra-pure water shield~\cite{ref:JUNO}.

To achieve the MH measurement, the precision of the reactor neutrino
spectroscopy has to be better than 1\% in energy scale and better than
3\% in the effective energy resolution~\cite{ref:energy_resolution}. These in turn pose a great challenge to the detector calibration. In Ref.~\cite{ref:calibration_strategy} we described the development of a comprehensive calibration strategy to address the
challenge, combining wisdom from past low energy neutrino experiments~\cite{ref:Daya_Bay_ACU, ref:KamLAND, ref:SNO, ref:Borexino, ref:SuperK, ref:Double_Chooz}. Several calibration hardware subsystems have been previously reported in Refs.~\cite{ref:CLS, ref:GTCS1, ref:GTCS2, ref:ROV}. In this paper, we discuss one of the key components of the
calibration system, the ACU, envisioned
as a primary tool to perform detector calibration on a regular basis. 

The paper is organized as follows. In Sec.~\ref{sec:Requirements}, a brief description
is given about the technical requirements of the ACU. The design of the ACU is
discussed in Sec.~\ref{sec:Design of the automatic calibration unit}. In Sec.~\ref{sec:Performance of the automatic calibration unit}, we discuss the performance of the as-built ACU, followed by the summary in Sec.~\ref{sec:Summary}.


\section{Requirements}
\label{sec:Requirements}

The main purpose of the ACU is to calibrate the energy scale and the
position non-uniformity along the central axis to a 0.1\% precision at each
position, and to calibrate the timing of the PMTs to sub-ns 
precision. As the most frequently operated calibration subsystem, the
technical requirements are:

\begin{enumerate}

\item It needs to be extremely reliable, and should be fully automated.
\item It should be able to deploy at least a gamma, or a neutron, or a
  UV laser light source, one at a time, into the CD along the central
  axis.
\item For flexibility, an auxiliary device (for example a temperature
  sensor) with standard connection interface can be easily deployed by
  the ACU.
\item The positioning accuracy needs to be controlled to less than 1~cm.
\item The ACU should be sealed with the CD to avoid an external radon
  leak. The total leakage requirement of the ACU is less than
  10$^{-5}$~mbar$\cdot$L$/$s~\cite{ref:leakage}.
\item Calibration sources induced background to the CD should be minimized.
\item Under no circumstances shall a calibration source drop into the
  CD or damage the CD.
\item The PMT readout should not be affected by electromagnetic noise from the ACU.
\item The ACU control system should be able to communicate with the trigger electronics. For example, if the trigger electronics detects a
  supernova burst, it should stop the ACU operation immediately.
\item The LS chemical compatibility and cleanliness: all mechanical
  parts directly or indirectly in contact with the LS must not influence 
  the properties of the LS.
\item Light should not leak into the CD through the ACU.

\end{enumerate}


\section{Design of the ACU}
\label{sec:Design of the automatic calibration unit}

\begin{figure*}[!htbp]
\centering
\includegraphics[width=0.8\textwidth]{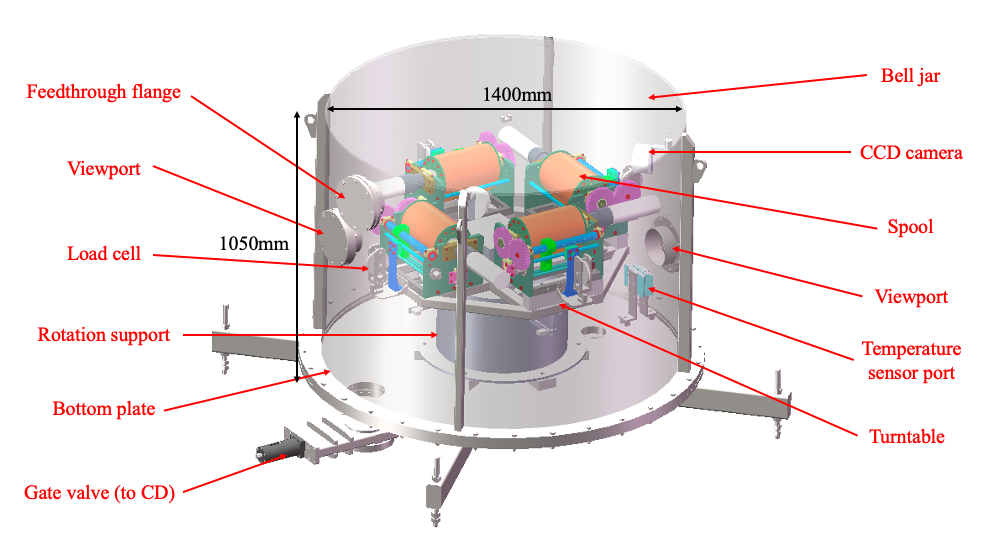}
\caption{An illustration of the design of the ACU.}
\label{ACU_overview}
\end{figure*}

An illustration of the design of the ACU is shown in
Fig.~\ref{ACU_overview}, similar to the calibration unit in the Daya Bay experiment~\cite{ref:Daya_Bay_ACU}. Details of the design features shall be discussed below.

\subsection{Material selection}
The JUNO LS is based on linear alkyl benzene (LAB)~\cite{ref:JUNO_LS}, 
which is a chemically mild organic liquid. The source and the
cable will be in direct contact with the liquid scintillator when
deployed into the CD. To avoid adverse long term effect to the LS,
especially the optical properties, we perform LS immersion tests on
various candidate samples, accelerated in an oven at
40~\textdegree{}C. The absorption spectrum of the liquid is measured before
and after the immersion. If the change of the deterioration rate is less than $10^{-4}$ (after the corrections of surface-to-volume ratios between the aging and experiment conditions, and the accelerated aging lifetime), the material is considered as a good material~\cite{ref:LS_compatibility}. Table~\ref{Material} shows a list of good materials, although some of the materials are only in
contact with the vapor of the LS in our system. All commercial components with unknown surfaces are enclosed within these good materials.

\begin{table}[htbp]
\centering
\caption{\label{Material}Materials used in the ACU.}
\smallskip
\begin{tabular}{ll}
  \hline
  Materials                              & components \\ \hline
  304 Stainless Steel (SS)                                    & most structural materials, bolts and nuts\\
  Aluminum extrusion                     & some structural materials\\
  Nickel                     & temperature sensor parts\\
  Ceramic & balls and bearings\\
  Polytetrafluoroethylene (PTFE) & electric cable jacket, source enclosure, spool \\
  Fluorinated Ethylene Propylene (FEP) & deployment cable jacket \\
  Polyurethane (PU) & electric cable jacket\\
  Acrylic & surface enclosure\\
  Viton & o-rings\\
  Ethylene Tetrafluoroethylene (ETFE) & optical fiber jacket\\
  Loctite epoxy & sealing glue\\
  Polyvinyl Chloride (PVC) & heat shrink tubing\\
  Polyacetal (POM) & spool gear\\
  Polyetheretherketone (PEEK) & bolts\\  \hline
\end{tabular}
\end{table}

To avoid induced radioactivity, the surface cleanliness of the ACU should follow the JUNO cleanliness 
protocol~\cite{ref:JUNO_cleanliness1, ref:JUNO_cleanliness2}. Radon emanation from material surface is another concern. Most of the 304 SS surfaces in the ACU are 
electropolished to a 1.6~$\mu$m roughness to reduce emanation. 
The ACU also contains N$_{2}$ flushing ports to bring out the radon as well as potential oxygen outgassing during the  operation, important for the radioactivity control and the light properties of the LS.
Moreover, the penetration hole of the ACU bottom plate
is sealed from the CD by a normally-closed gate valve.

\subsection{Mechanical design}

\begin{figure*}[!htbp]
\subfigure[]{
\begin{minipage}[t]{1\linewidth}
\centering
\includegraphics[width=0.75\textwidth]{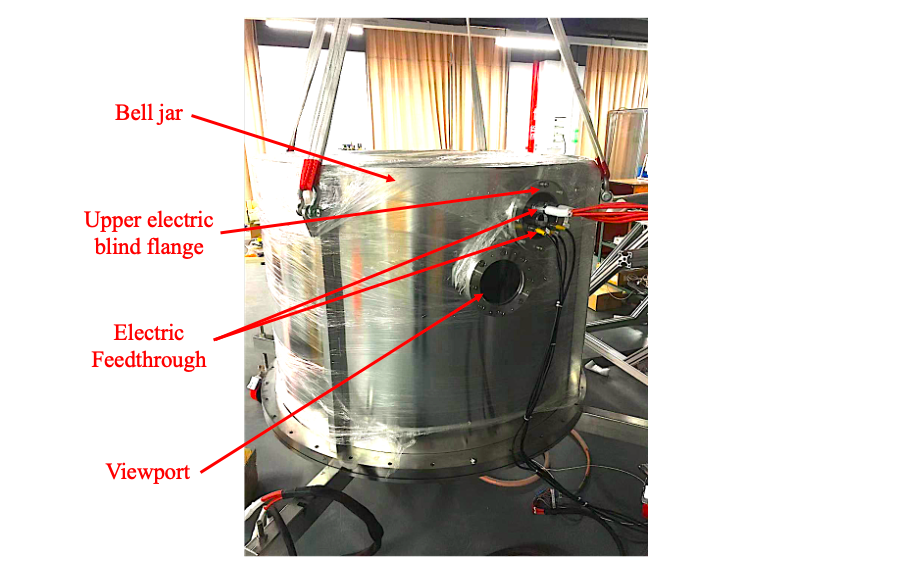}
\label{ACU_bell_jar}
\end{minipage}
}

\subfigure[]{
\begin{minipage}[t]{1\linewidth}
\centering
\includegraphics[width=0.8\textwidth]{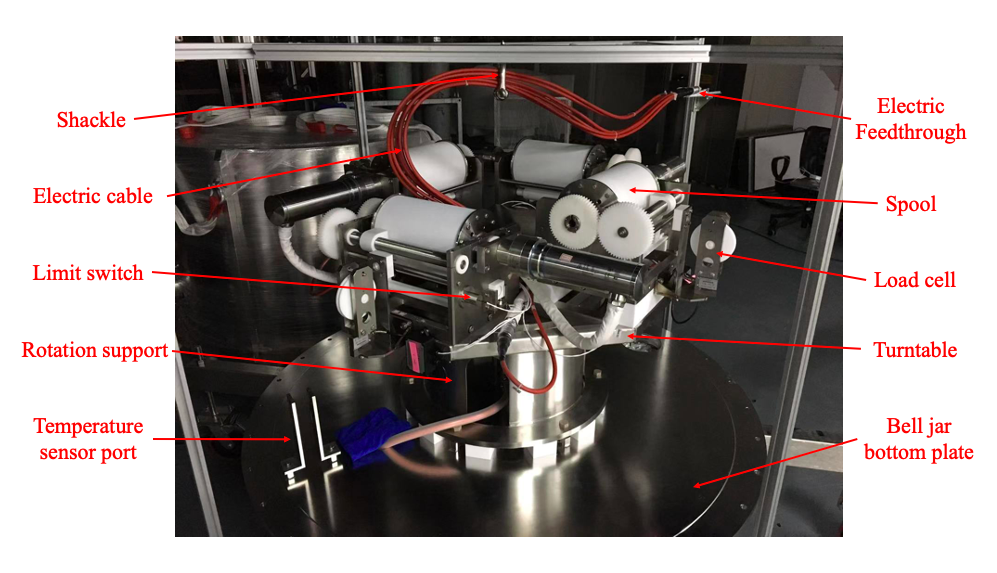}
\label{ACU_major_modules}
\end{minipage}
}
\caption{(a) Photo of the ACU bell jar; (b) All the ACU components mounted onto the bottom plate.}
\end{figure*}

The ACU is enclosed in a stainless steel vacuum sealed bell jar with
1400~mm inner diameter and 1050~mm height (Fig.~\ref{ACU_bell_jar}). Two
viewports (normally blanked off to avoid light leaks) and one electrical feedthrough flange are on the side wall
of the bell jar. The bottom plate of the ACU is 18~mm thick with one electrical feedthrough flange and a 150~mm diameter hole aligning with the central axis of the CD, 620~mm offset from the center of the bottom plate, 
through which the source can be lowered
into the CD. During the regular non-calibration data taking, a gate valve seals the hole from the underside of the bottom plate.

A picture of the complete ACU without the bell jar is shown in Fig.~\ref{ACU_major_modules}, we will discuss all of the components in turn.

\subsubsection{Turntable}
\begin{figure*}[!htbp]
\centering
\includegraphics[width=0.75\textwidth]{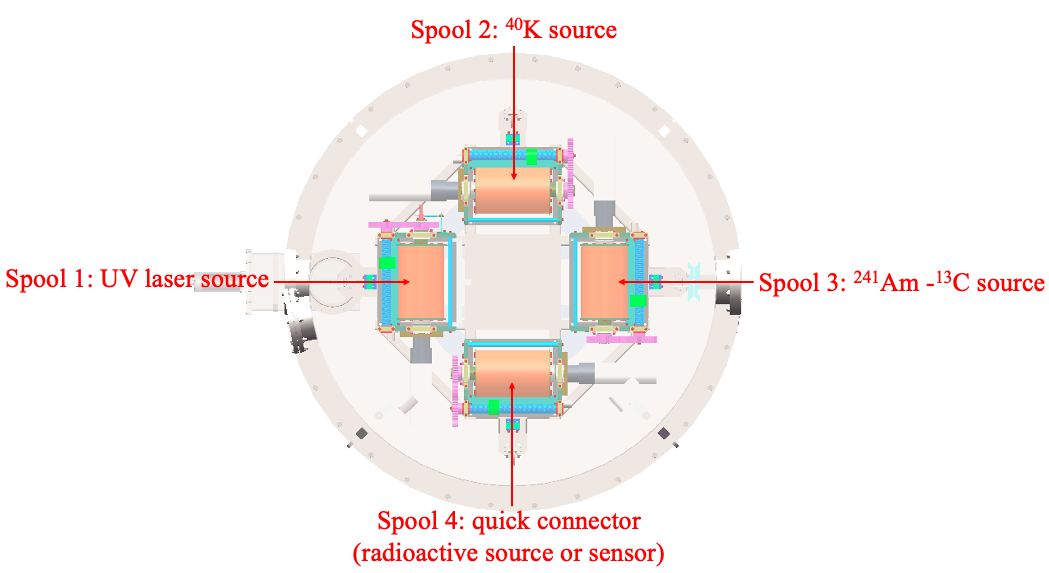}
\caption{Top view of the ACU turntable.}
\label{ACU_top_view}
\end{figure*}

The body inside the ACU is a steel turntable including four mechanically independent motor/spool mechanisms. Each spool is able to deploy a source (a UV fiber with a diffuser ball, a radioactive source, or some auxiliary device which can be attached to the standard source connector) into the CD
along the central axis. The four spools are installed with azimuthal
separation of 90$^{\circ}$, as shown in Fig.~\ref{ACU_top_view}.

The turntable is capable of turning clockwise or counter-clockwise by
a servo motor installed at the center of its supporting structure. The
center of each source is radially displaced from the center of the
turntable by 620 mm. Once it is aligned with the 150~mm 
hole at the bottom plate, the source can be deployed into the CD.

\begin{figure*}[!htbp]
\centering
\includegraphics[width=0.75\textwidth]{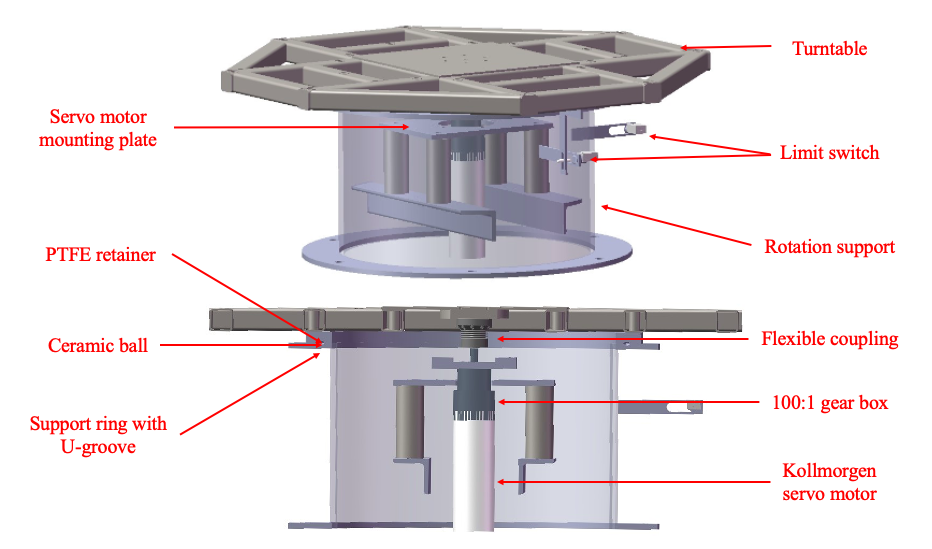}
\caption{Drawings of the turntable.}
\label{ACU_turntable}
\end{figure*}

Mechanical details of the turntable are illustrated in
Fig.~\ref{ACU_turntable}. A stainless steel rotation support is installed at the center of the bottom
plate. The top surface of the rotation support is a U-shaped groove, hosting
40 12.3~mm ceramic balls, evenly distributed by a PTFE
retainer. This mechanism serves both as a support to the turntable and the bearing for its rotation. An IP69K watertight servo motor (Kollmorgen
AKMH-CNT2GE5K) coupled to a 100:1 gear box (Thomson
AQT060-100-0-MMR-724) is installed at the center of the rotation
support. The shaft of the gear box couples to the turntable via a
flexible bellow coupling (R+W BK8 series). In order to limit the range of the rotation, two push-button-style stainless steel limit switches
(Schurter) are installed on the rotation support to define the limits.

\subsubsection{Spool for the radioactive source}
\begin{figure*}[!htbp]
\centering
\includegraphics[width=0.75\textwidth]{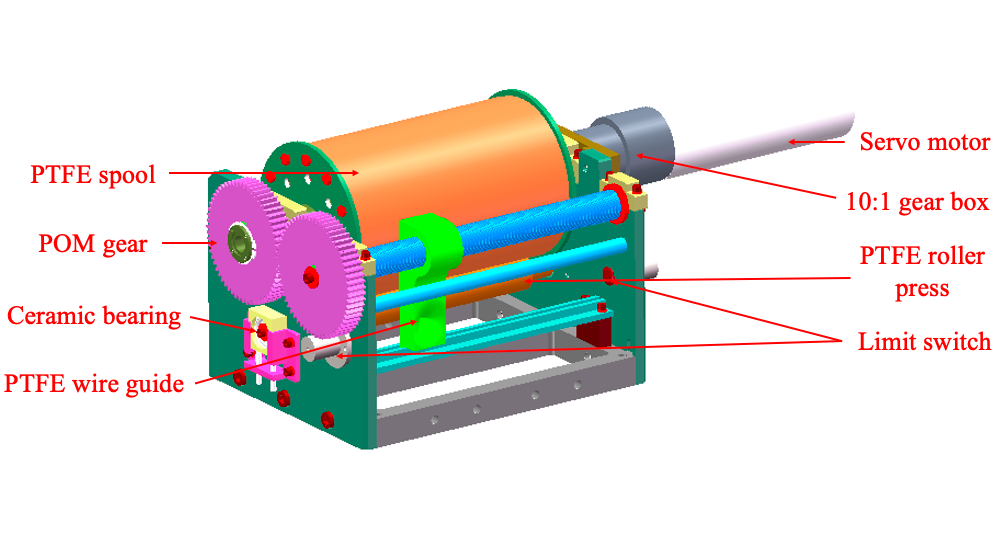}
\caption{Drawing of the radioactive source spool.}
\label{ACU_spool}
\end{figure*}

\begin{figure*}[!htbp]
\centering
\includegraphics[width=0.85\textwidth]{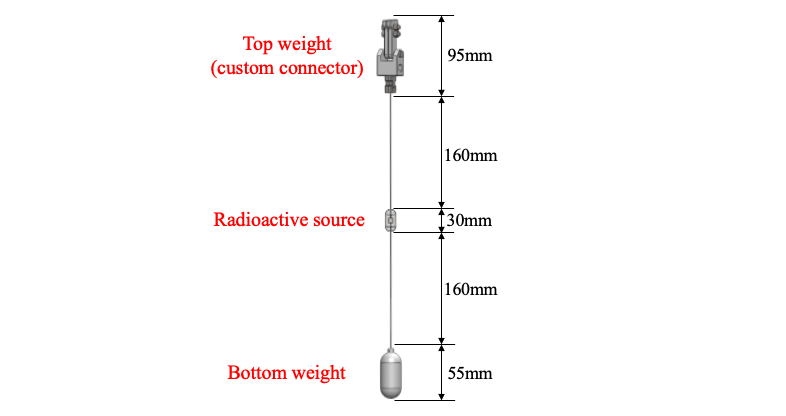}
\caption{A schematic diagram of a radioactive source assembly. A custom connector serves as a top weight. The 6$\times$6~$mm^{2}$ SS radioactive source capsule is enclosed in a PTFE enclosure. A Nickel block enclosed by a PTFE enclosure serves as the bottom weight.}
\label{source_assembly}
\end{figure*}

Fig.~\ref{ACU_spool} illustrates details of a spool for radioactive
source deployment. It consists of an IP69K watertight servo motor
(Kollmorgen AKMH-CNT2GE5K) coupled to a 10:1 gear box (Thomson
AQT060-010-0-MMR-724) and a 160~mm diameter PTFE deployment spool with
helical grooves. An FEP-coated flexible 1~mm diameter stainless steel cable is wound in the grooves. The open end of the cable runs through a PTFE pulley, then gets attached to a radioactive source (Fig.~\ref{source_assembly}). When
the servo motor drives the spool to unwind the cable, the source can be deployed under gravity into the CD. To minimize the risk of the cable exiting a groove, a PTFE cable guide driven by two POM gears moves synchronously with the cable. In addition, 
a PTFE roller press keeps the cable in the grooves. The motion limits are set by two push-button-style stainless steel limit switches (Schurter).
 
\subsubsection{Spool for the UV fiber}
The design of spool for the UV fiber source is slightly
different, as we need to deploy a multi-mode optical
fiber to transmit the UV laser light into the
CD. Details of this spool can be found in Ref.~\cite{ref:laser}. 
An optical fiber rotary joint (MOOG FO228)
is connected with the shaft of the spool to link the optical fiber placed outside
the ACU bell jar to the fiber rotating with the spool. 

\subsubsection{Spool for the temperature sensor}
\begin{figure*}[!htbp]
\centering
\includegraphics[width=0.75\textwidth]{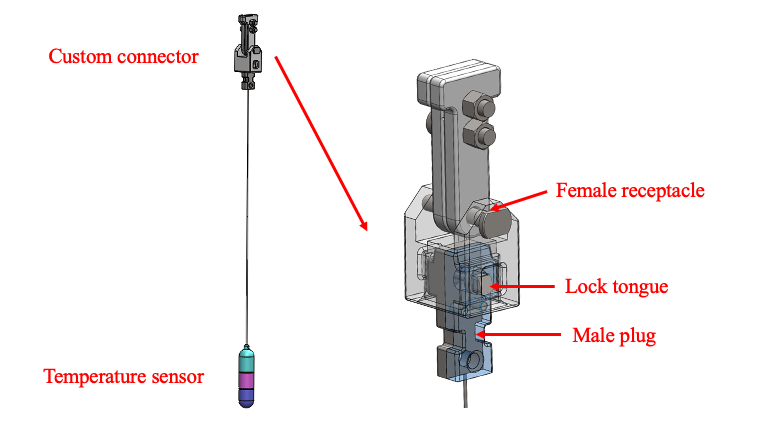} 
\caption{A schematic diagram of temperature assembly (left) and
  the custom connector (right).}
\label{temperature_assembly}
\end{figure*}

The fourth spool is envisioned to be flexible. The open end of the cable
is attached to a custom connector (see Fig.~\ref{temperature_assembly}), which is able to attach either an auxiliary sensor or a radioactive source. The
quick connector consists of a female receptacle and a removable male plug. The plug can be inserted straight into the receptacle. To detach the plug, one has to use a pin tool to push in both tongues then pull to separate. 
This simple manual operation (although envisioned to be very rare) can be performed by opening the ACU viewport. To be safe, the gate valve should be shut, so this action can be performed even when the JUNO is fully operational. 

\begin{figure*}[!htbp]
\centering
\includegraphics[width=0.75\textwidth]{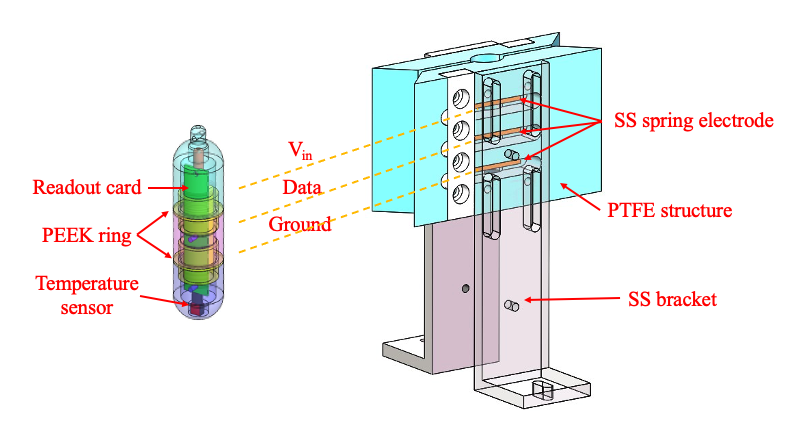}
\caption{Drawing of the temperature sensor and its readout port.}
\label{ACU_TS_and_TSport}
\end{figure*}

We designed a watertight wireless temperature sensor to facilitate the measurement of the LS temperature along the central axis. The LS temperature affects the neutron capture time. It also affects the speed of the sound, a key input for the ultrasonic source positioning system~\cite{ref:USS}.
As shown on the left side of Fig.~\ref{ACU_TS_and_TSport}, the temperature sensor consists of a three-part
shell, each serving as an electrode (from top to bottom: V$_{\rm in}$,
Data, and Ground), insulated by PEEK rings. The temperature sensor
(INNOVATIVE SENSOR TECHNOLOGY TSIC506, 0.1~$^\circ$C), its digital
readout card, and a rechargeable battery are sealed inside the shell,
with the sensor head in tight thermal contact with the lower shell,
made with pure Nickel for better thermal conductivity. The temperature data will be stored on the readout board with timestamps during the
deployment. Shown on the right side of Fig.~\ref{ACU_TS_and_TSport} is the readout port mounted on the
bottom plate, for reading the temperature data and recharging the
battery. 
The port has three pairs of stainless steel
spring electrodes to make electrical contact with corresponding pieces on
the temperature shell. To read the data, the temperature sensor will
be moved directly below the port, and get pulled upward in place. When the wire gets retracted further upward, 
the sensor gets released from the port to the parking position.

\subsubsection{Load cell}
\begin{figure*}[!htbp]
\centering
\includegraphics[width=0.75\textwidth]{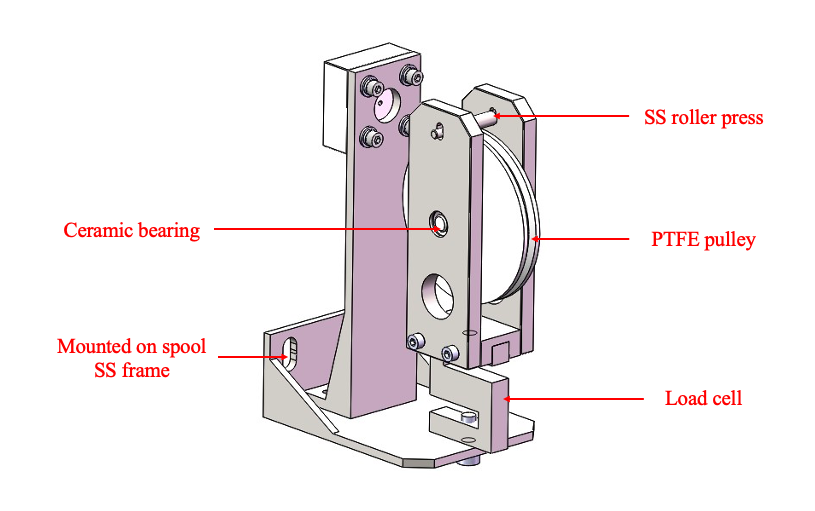}
\caption{Drawing of the load cell.}
\label{ACU_load_cell}
\end{figure*}

A PTFE pulley with a ceramic bearing serves as a guide wheel for the
calibration cables. A stainless steel S-shape load cell (Longlv LLBLS-I) with a full range of 5 kg and an accuracy 1.5~g, is mounted below the pulley to constantly monitor the
tension in the cable (Fig.~\ref{ACU_load_cell}). To set the scale, the weight of the cable is approximately 3~g/m, and that of the entire source assembly is about 200~g. Too much tension may cause damage to the cable or joints, and a loss of tension could result in the cable slipping out of in the grooves of spool.

Because the raw analog outputs of the load cell are of the order of mV, to
avoid noise pickup, a digital transducer (SMOWO RW-ST01D 2.1) is used to digitize the signal into RS485 signals before
transmission.

\subsubsection{Cameras}
\begin{figure*}[!htbp]
\centering
\includegraphics[width=0.75\textwidth]{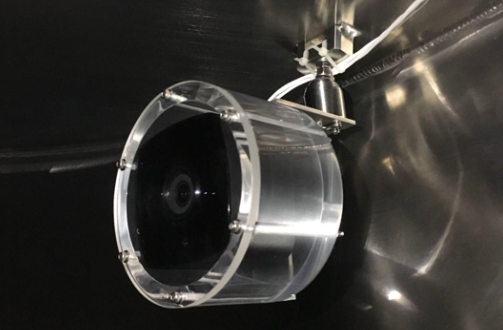} 
\caption{Picture of the CCD camera enclosed in an acrylic cylinder.}
\label{ACU_CCD}
\end{figure*}

\begin{figure*}[!htbp]
\subfigure[]{
\begin{minipage}[t]{1\linewidth}
\centering
\includegraphics[width=0.75\textwidth]{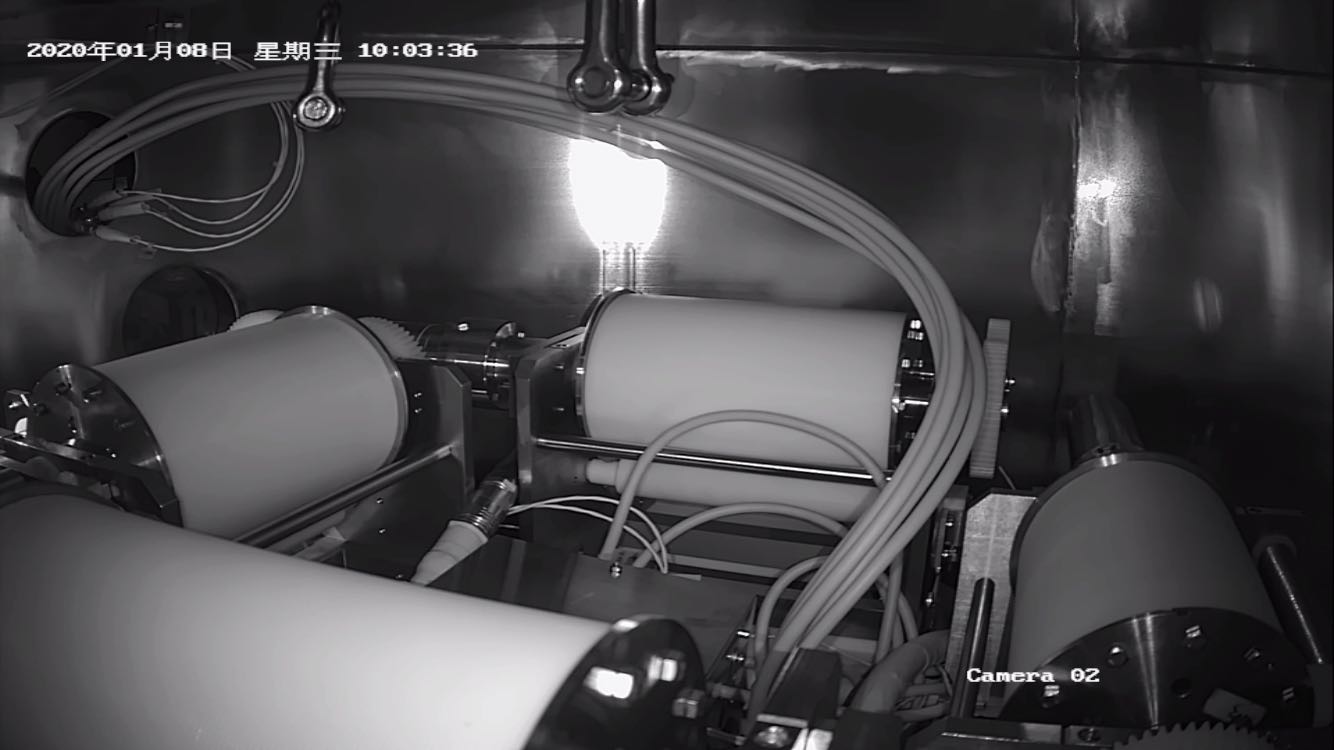}
\label{ACU_CCD_output_1}
\end{minipage}
}

\subfigure[]{
\begin{minipage}[t]{1\linewidth}
\centering
\includegraphics[width=0.75\textwidth]{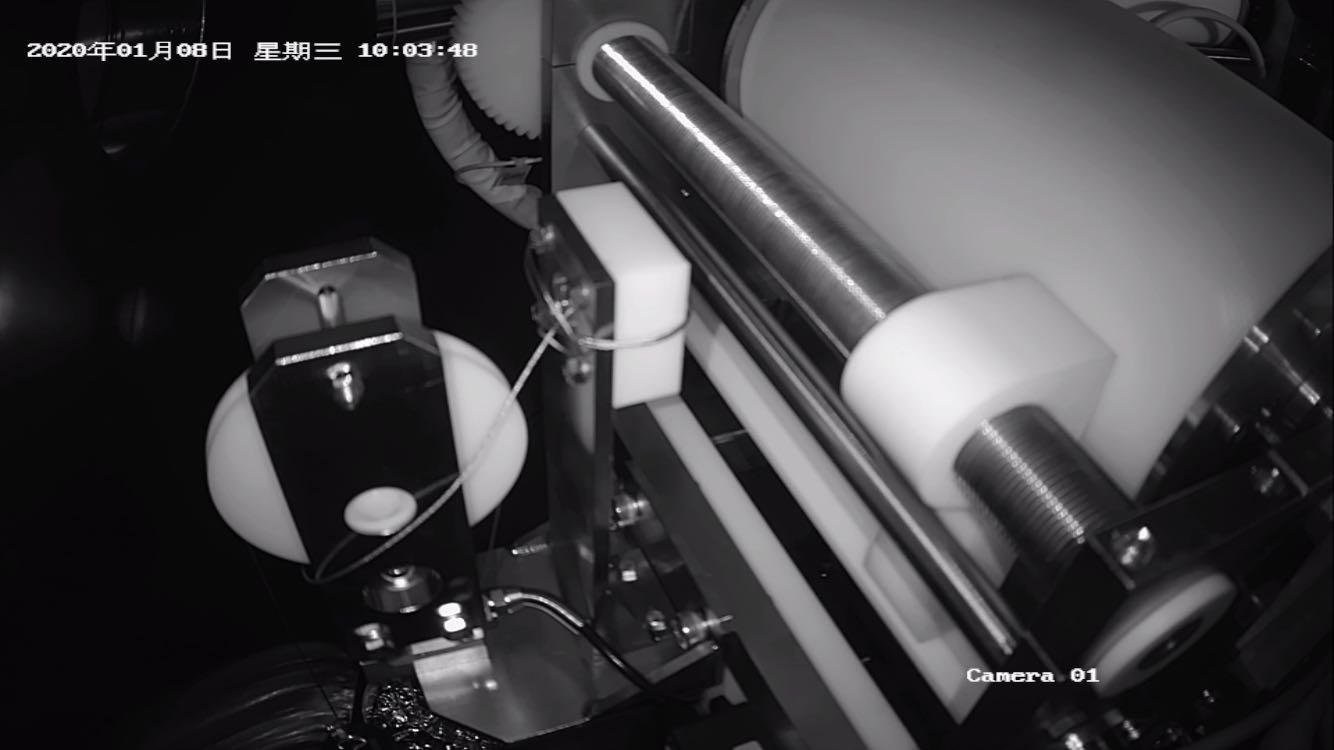}
\label{ACU_CCD_output_2}
\end{minipage}
}
\caption{Images of the interior of the ACU taken by the two CCD cameras.}
\end{figure*}

In order to remotely monitor the components inside the ACU, two CCD
cameras (HIKVISION DS-2CE16D1T-IT3F) with infrared LED lighting are
installed on the inner barrel of the bell jar. Fig.~\ref{ACU_CCD} is a picture of the
CCD camera in a custom acrylic enclosure. 
The lighting will be
turned off during normal data taking, so it will not affect the CD PMTs. Two images taken by the cameras 
are shown in Figs.~\ref{ACU_CCD_output_1} and \ref{ACU_CCD_output_2} when the ACU is kept in the darkness.

\subsubsection{Interface to the CD}
\begin{figure*}[!htbp]
\centering
\includegraphics[width=0.75\textwidth]{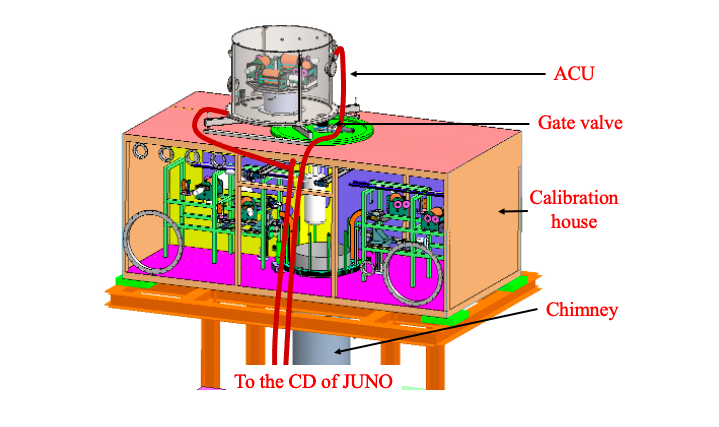} 
\caption{A schematic diagram of the ACU interface to the CD.}
\label{ACU_interface}
\end{figure*}

Fig.~\ref{ACU_interface} is a schematic diagram of the ACU interface
to the CD. The ACU is installed on the top of the calibration house, where other calibration devices are located. The 150 mm diameter gate valve, through which the sources go in and out of the ACU, is sealed to calibration house through a flexible stainless steel bellow. The bottom of the calibration house seals to 
a 7.85~m long stainless steel chimney of the CD. Unlike the ACU system in the Daya Bay experiment which sat right on
top of the neutrino detector under water~\cite{ref:Daya_Bay_ACU}, the JUNO ACU is installed in the experimental hall, separated from the CD by the water shielding. The nominal rate for each source is set to be around 100 gamma or neutron emissions per second~\cite{ref:calibration_strategy}. Their contribution to the CD rate is limited to less than 1 mHz, according to the simulation. Therefore, no additional shielding inside the ACU is needed.

The bell jar seals to the bottom plate by a double
o-ring seal, so that leak check can be performed by pumping the volume between the two o-rings. A Helium sniffer leakage detection technique was used off-site of JUNO in the quality assurance procedure. The upper limit of leakage rate was measured to be 3 $\times$ 10$^{-6}$~mbar$\cdot$L$/$s.

\subsection{Electronics and wiring}
\subsubsection{Internal wiring}
\begin{figure*}[!htbp]
\centering
\includegraphics[width=0.75\textwidth]{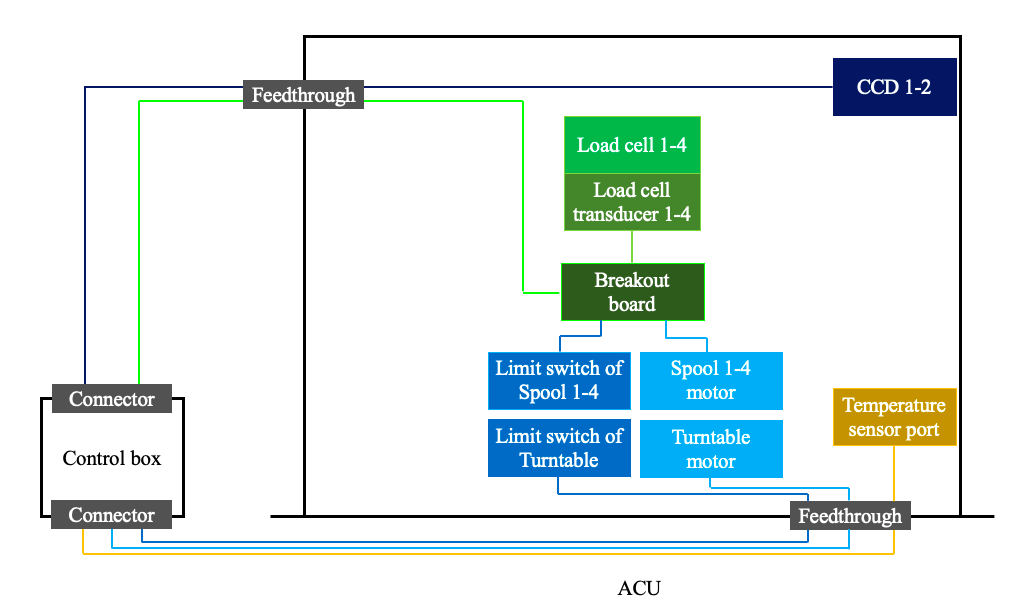} 
\caption{A schematic diagram of the ACU wiring scheme.}
\label{ACU_wiring}
\end{figure*}

A schematic diagram of the ACU's overall
wiring scheme is shown in Fig.~\ref{ACU_wiring}.

Two 48-pin feedthroughs (LEMO SGJ4B348CLMPV) and four isolated BNC
feedthroughs (MPF A0466-4-W) are mounted on an ISO-160 vacuum
flange (upper electric flange). The two 48-pin feedthroughs are used for
connections of the components on the
turntable. Two of the BNC connections carry the 12~V power supply
for each CCD cameras, and the other two carry the image signals.
The feedbacks of the servo motors, and the load cell power supplies and signals get collected by a breakout board
located inside a shielded stainless steel box on the
turntable. To avoid crosstalk between the power and feedback of the
servo motors, their power lines and limit switches get consolidated by
the other breakout board, shielded from the first one. Through multiple twist-pair cables, each breakout board connects to a 48-pin female connectors (LEMO FGG4B348CLAD15 and FGJ4B348CLAD15), mating to
the feedthroughs on the upper electric flange. All twisted-pair cables go upward through a rotatable shackle from the center of the bell jar dome, then run horizontally to connect to the
upper electric flange. This ensures that cables will not run into
other components while the turntable is rotating.

One 30-pin feedthrough (LEMO SGJ4B330CLMPV) is mounted on an ISO-100 vacuum
flange (lower electric flange), connecting to the turntable servo motor, its limit switches, and the temperature sensor port.

\subsubsection{External wiring}
Outside the ACU, the cables ($\sim$10 m) go all the way from the upper and
lower electric flanges to the control box located on the platform below
the calibration house (Fig.~\ref{ACU_interface}). 
The wiring is also indicated in Fig.~\ref{ACU_wiring}.

\begin{figure*}[!htbp]
\subfigure[]{
\begin{minipage}[t]{1\linewidth}
\centering
\includegraphics[width=0.75\textwidth]{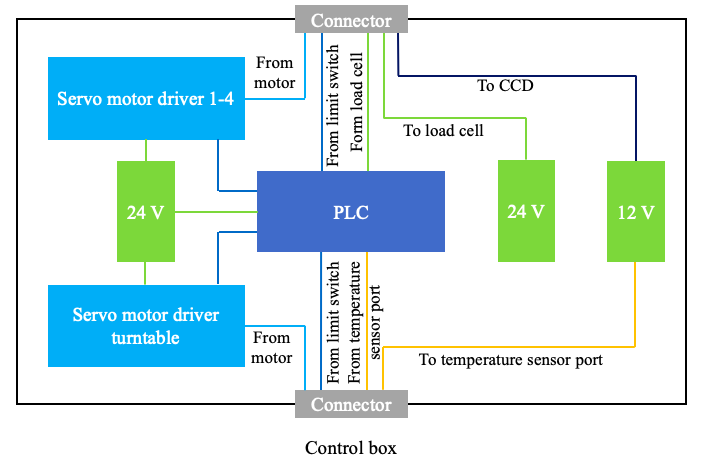}
\label{ACU_control_box_diagram}
\end{minipage}
}

\subfigure[]{
\begin{minipage}[t]{1\linewidth}
\centering
\includegraphics[width=0.75\textwidth]{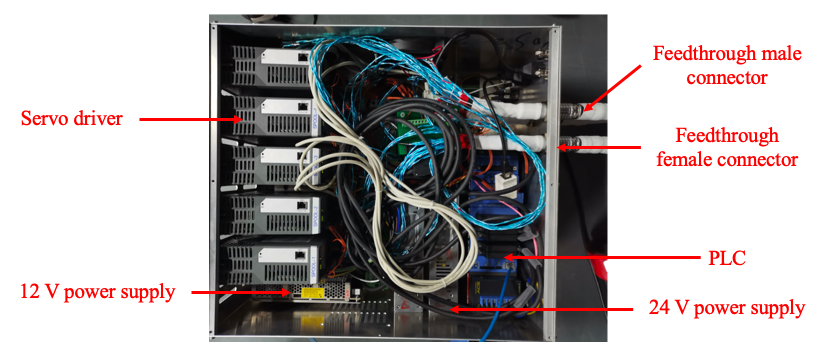}
\label{ACU_control_box}
\end{minipage}
}
\caption{A diagram (a) and a picture (b) of the ACU control box.}
\end{figure*}

A diagram and picture of the ACU control box are shown in Figs.~\ref{ACU_control_box_diagram} and \ref{ACU_control_box}. It
consists of five servo drivers for the spool and turntable servo
motors, a programmable logic controller (PLC) (YOKOGAWA), one 12~V
power supply for temperature sensor port and CCD cameras, and two 24~V 
power supplies
for load cell transducers and servo drivers. The motion control, data acquisition, and communications are all controlled by
the PLC. To avoid servo motors introducing noises in the CD, relays are implemented in the control box to remotely disable the servo
drivers, etc., when the sources are not moving. 

The power line of the control box is drawn from the main AC 220 V power from the electronics room to allow single point grounding.  The
PLC in the control box is connected via an Ethernet cable to a Windows~10 computer located in the electronics room, serving as a state monitor.

\subsubsection{Groundings}
\begin{figure*}[!htbp]
\centering
\includegraphics[width=0.75\textwidth]{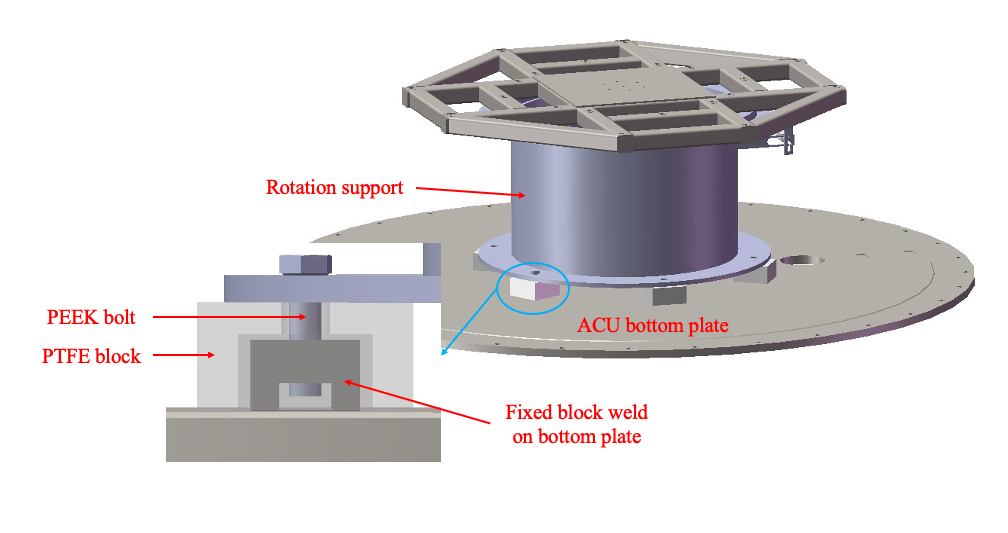} 
\caption{The ACU electronics insulation plan.}
\label{ACU_insulation}
\end{figure*}

JUNO has two independent power and ground systems: so-called the "clean" and "dirty". The PMT and its electronics system
run on the clean power and clean ground, whereas most of the utilities
run on the dirty power and safety ground. Since the ACU uses motors, by
default, the ACU will be connected to the dirty power. On the other
hand, most CD structure will be on the clean ground, therefore it is imperative that the electronics
components inside the ACU be completely insulated from the bottom
plate. To achieve this, the rotation support is insulated from the bottom plate by PTFE spacers and PEEK
bolts, as illustrated in Fig.~\ref{ACU_insulation}.

\subsection{Design of the control software}
\begin{figure*}[!htbp]
\centering
\includegraphics[width=0.75\textwidth]{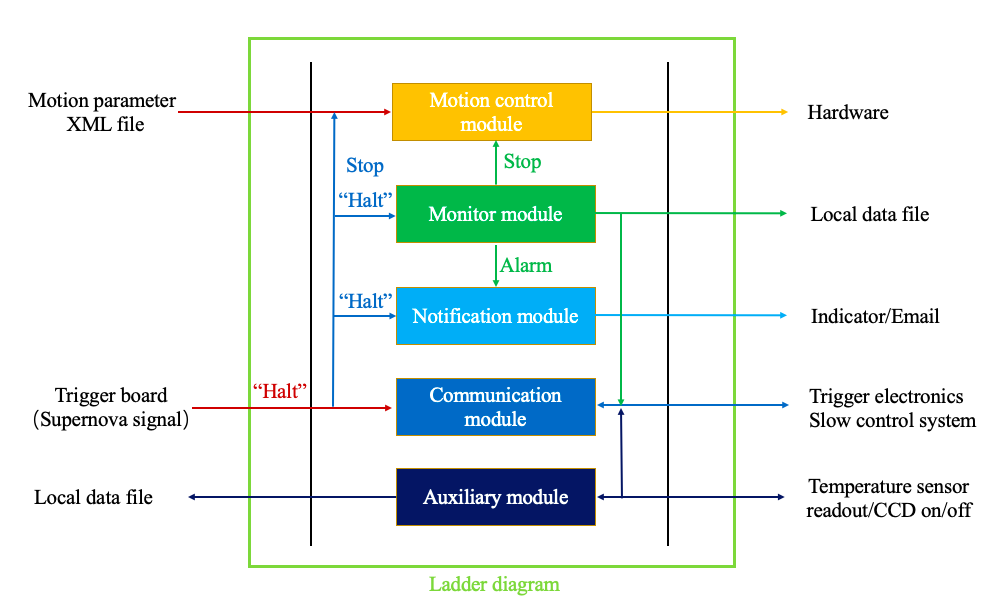} 
\caption{The ladder diagram of the ACU control software.}
\label{ACU_software_structure}
\end{figure*}

The design of the ACU control software follows two guidelines: 1) safety, i.e. it should be reliable to interrupt motions immediately when abnormal conditions are detected; and 2) automation, i.e. the calibration sequence is carried out completely without human
intervention. The control software is developed in a ladder diagram (the programming language of
the PLC), illustrated in
Fig.~\ref{ACU_software_structure}, which consists of five application
modules: motion control, monitoring, notification, 
communication, and auxiliary modules. An entire execution cycle was found to take less than 0.1~s. The motion control module controls the five axes (1-4 for the spools and 5 for the turntable) of motion. In the standard operation, it reads an XML script which
specifies the sequence and detailed parameters of operations to be
performed. The monitor module reads data from all sensors. When the values of these sensor exceed prescribed values, the motion control module will stop
all motions and the corresponding alarms will appear.
The alarms will be broadcast by the notification module. The auxiliary module controls the switches of the CCD
cameras and communicates with the temperature sensor port. The communication module talks to the slow control system and the trigger electronics, particularly during the calibration data taking. In addition, when the trigger electronics detects something interesting in realtime (e.g. supernova neutrino burst), the trigger electronics handshakes with the communication module which issues halt signals to all motors. 

\subsection{Safety considerations}
Robustness of the ACU against possible failures is of ultimate importance. Similar to the design of another calibration subsystem in Ref.~\cite{ref:CLS}, the safety considerations implemented in the ACU control are summarized below. 

\begin{enumerate}
\item All motions are relatively slow. The turntable rotates at a
  speed of 2.25$^{\circ}$$/$s, and the sources are deployed at a speed
  of 15 mm$/$s. This not only gives reaction time if something goes wrong, but also improves the longevity of servo motor operations.
\item The servo motors used in the ACU are equipped with electromagnetic
  brakes which will lock the motors during power failure.
\item The rotation angle of the turntable is limited to 275$^{\circ}$
  by two limit switches to avoid over-twisting of electric cables and the laser fiber. The deployment of the spool is also limited by two
  limit switches to avoid motion beyond allowed range.
\item Each radioactive source assembly contains a top weight and
  bottom weight above and below the source to maintain the tension in the cable. The double-weight ensures 
  sufficient tension even if the bottom one is touching some obstacle from below.
  \item The tension in the deployment cable is constantly being monitored by the load cell to avoid too much or too little tension when the source is stuck or bottomed out.
 \item The source enclosure and weights are round-headed to avoid the source getting stuck or damaging the surface that it scrapes on.
\item The breaking strength for the FEP-coated stainless steel cable
  is about 60 kg. In order to avoid the cable breakage due to the load
  cell failure, the torque of the servo motor is limited to provide a
  maximum tension of less than 10~N in the cable. The control
  software will monitor the current of the servo motor in real time,
  and shut down the motor when there is excessive current. The same
  current interlock is also implemented in the turntable motor. 
  \item The PLC has an internal battery, to allow data to continue being buffered during a power failure.
\end{enumerate}

\section{Performance of the ACU}
\label{sec:Performance of the automatic calibration unit}

A complete ACU has been constructed according to the above design. Rigorous functionality
tests have been carried out together with the control software, yielding satisfactory performance in all aspects. 

\subsection{Positioning accuracy}
\begin{figure*}[!htbp]
\centering
\includegraphics[width=0.75\textwidth]{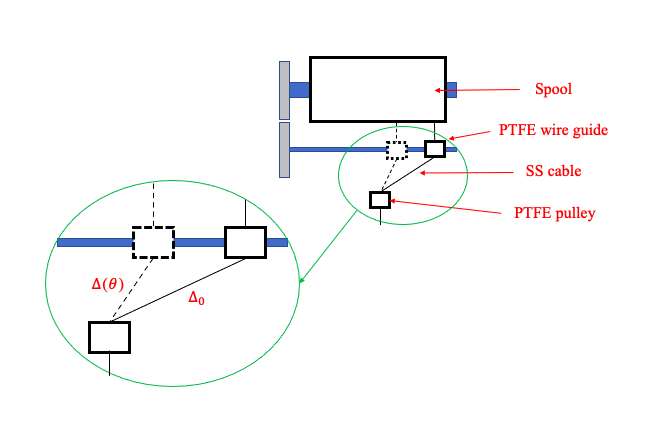} 
\caption{An illustration of $\Delta_0$ and $\Delta(\theta)$.}
\label{spool_guide}
\end{figure*}

\begin{figure*}[!htbp]
\centering
\includegraphics[width=0.75\textwidth]{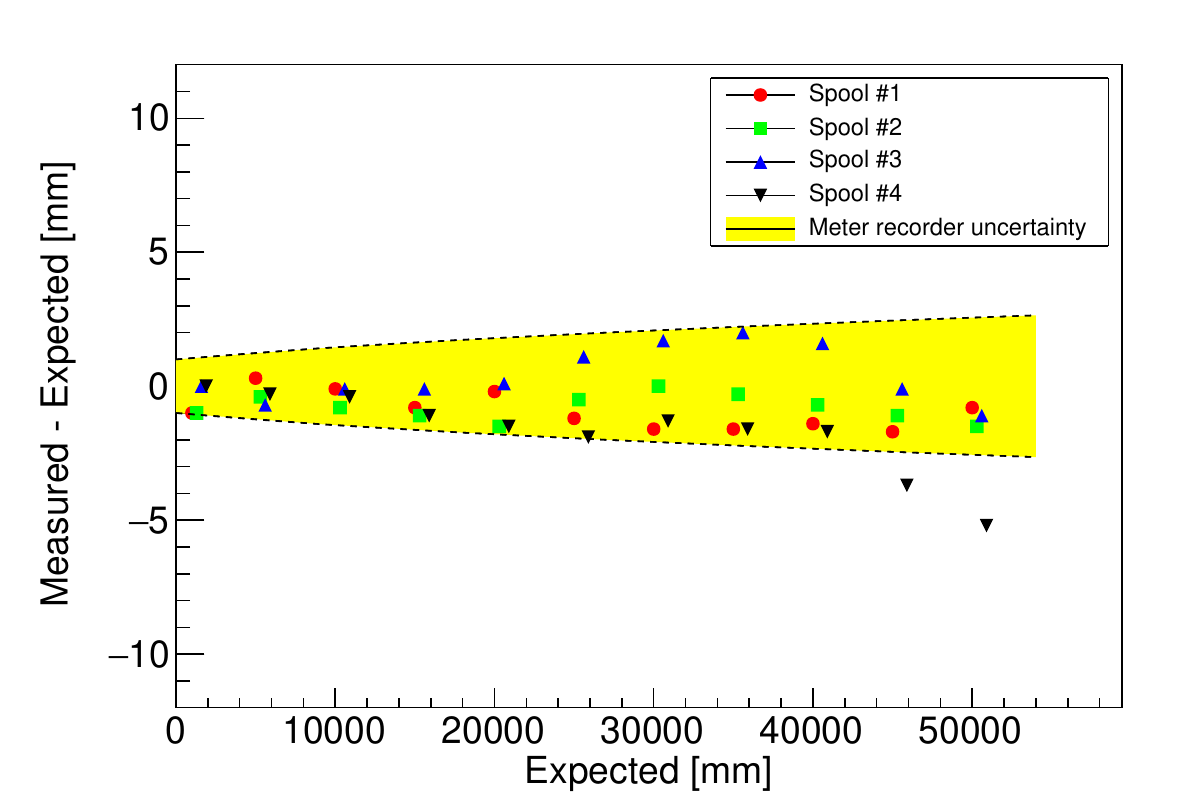} 
\caption{Biases in deployed cable length (measured-expected) vs. expected distance for all spools, with color coding indicated in the legend. The
yellow band indicates the expected uncertainty of the length measured by the meter recorder.
}
\label{spool_positioning_uncertainty}
\end{figure*}

The source deployment position is determined based on two assumptions.
First, the cable is within the elastic limit under the typical load of a few hundred grams, so the fractional elongation under gravity is a constant. This assumption has been validated by having two cables of 24 and 35~m hung vertically with a weight varying from 0 to 300~g repeatedly.
Good elasticity has been observed, and under the 200 g source load, the elongation was measured to be $1.0\times10^{-4}$ at 35~m. Even at the maximum length, the elongation is limited to be 5~mm. This can be conservatively taken as an uncertainty. Furthermore, the potential curls in the deployment cable might introduce additional positioning uncertainty in the horizontal plane. After the detector is assembled, we will use the four spools to deploy different sources to the same nominal location, and use the auxiliary CCD cameras~\cite{ref:CCD} (4 mm relative positioning accuracy) to check the position repeatability.
Second, the spool diameter is a constant which relates the unwound cable length and the spool rotation as
\begin{equation}
\label{eq:length}
    L = \theta/2 \times D + \Delta_0 - \Delta(\theta)\,, 
\end{equation}
in which $\theta$ is the spool rotation in radian, D is 
the effective diameter of the spool, and $\Delta_0$ and $\Delta(\theta)$ are illustrated in Fig.~\ref{spool_guide}, calculable based on simple trigonometry. 
Systematic uncertainties can arise due to machining precision of the grooves and how well the cable fits inside the grooves. To study this, for
each spool, we unwind and stop the spool every five cycles, and record the cable length using a custom length recorder, which was calibrated against a 9~m tape measure to a precision of 1~mm. Then using Eqn.~\ref{eq:length}, the spool diameter $D$ is fitted as a floating parameter. To test the positional bias, we then unwind the spool again at 1, 5, 10, 15, 20, 25, 30, 35, 40, 45, and 50~m according to Eqn.~\ref{eq:length}, 
then measure directly the cable length using the length recorder. The differences between the measurements and 
expectations vs. distance are summarized in Fig.~\ref{spool_positioning_uncertainty} for 
the four spools. The yellow 
band indicates the accumulated systematic uncertainty of 
the length recorder. Since most of the 
data points are inside the band, and the largest difference is 5.2~mm (spool 4) at 50~m, we conclude that the uncertainty due to the spool rotation is controlled to be less than 5~mm. In combination with an independent uncertainty due to cable elongation (5~mm), the total vertical position uncertainty is estimated to be 7~mm.

\begin{figure*}[!htbp]
\centering
\includegraphics[width=0.75\textwidth]{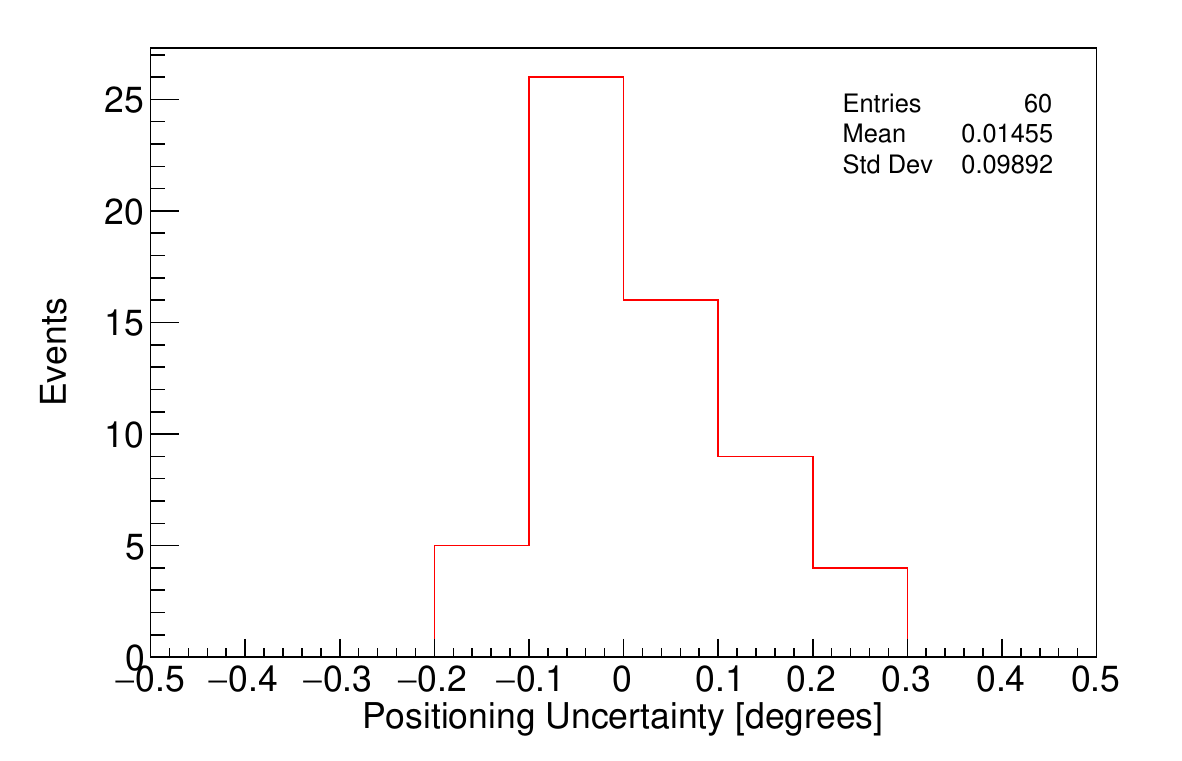} 
\caption{The measured angular uncertainty of the turntable. }
\label{turntable_positioning_uncertainty}
\end{figure*}

Another ingredient of the positional precision is in the 
horizontal plane. To deploy a given source, the turntable needs to be rotated to a specified
position. Although the precision of the servo motor and gear rotation is very high, the zero position of the turntable bears non-negligible uncertainty due to the spring actuation of the limit switch. To study this, the turntable is continuously moved back and forth between the zero and 270$^{\circ}$ 60 times, each time resetting zero by hitting the limit switch, then stopping at 270$^{\circ}$. The angle between the stopped position and a reference plate is measured by a dial indicator, shown in Fig.~\ref{turntable_positioning_uncertainty}. The standard deviation of the angle, translated to a horizontal position uncertainty at a radius of 620~mm (the eccentricity of the deployment hole) is 1.1~mm.

\subsection{Load cell sensitivity}
\begin{figure*}[!htbp]
\centering
\includegraphics[width=0.75\textwidth]{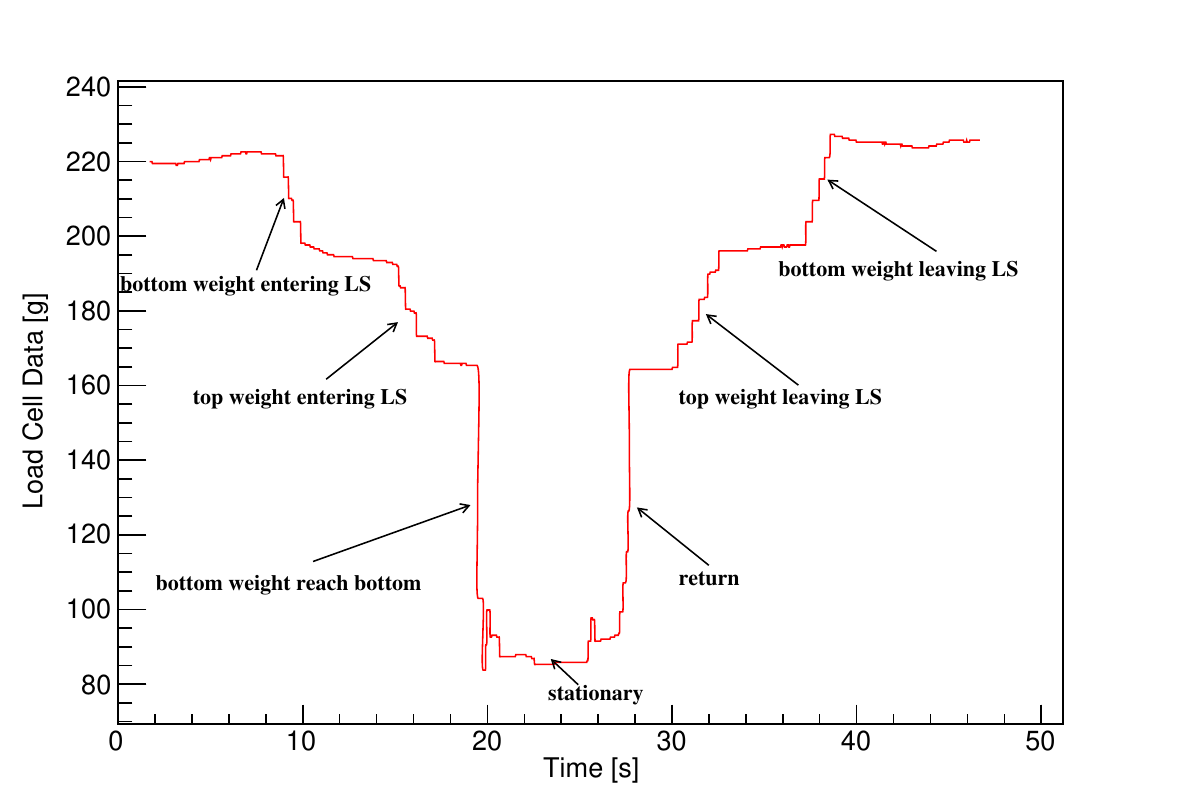} 
\caption{Load cell reading for a test deployment of a dummy source into and out of the LS.}
\label{ACU_load_cell_test}
\end{figure*}

The load cell is tested with a dummy source deployed from the air into the LS, with its reading depicted in 
Fig.~\ref{ACU_load_cell_test}. The change of the tension
due to different buoyancy conditions as well as a "bottom-out" situation can be clearly observed. After the source completely
leaves the LS, the tension has another slight decrease as the LS initially attached to the source drips back into the liquid, demonstrating the sensitivity of the load cell.

\subsection{Exercised safety features}
We have also made repeated tests on all safety measures in the control software. The test results are summarized in Table~\ref{software_test}.

\begin{table}[]
\centering
\caption{\label{software_test} The ACU control software safety measure tests.}
\smallskip
\begin{tabular}{lll}
\hline
test item               & test method      & test result                                   \\ \hline
motion interlock        & $>$ 1 servo motor running        & all motions stopped, and alarm issued \\
turntable limit switch  & triggered by hand          & same as above \\
spool limit switch      & triggered by hand         & same as above \\
load cell               & out of range & same as above \\
servo motor current     & out of range & same as above \\
power outage            & power cycle by hand & \begin{tabular}[c]{@{}l@{}} all motion stopped, electromagnetic\\ brake on, and all data recovered after\\ power cycle\end{tabular}\\
\hline
\end{tabular}
\end{table}


\section{Summary}
\label{sec:Summary}
In this paper we described the design of the automatic calibration unit for the JUNO experiment, which is capable of deploying multiple radioactive sources, a UV laser source, or an auxiliary sensor along the central axis of the JUNO CD with positional precision better than 1~cm. Long term reliability has been built into the hardware and software design. The production version of the ACU is constructed with its performance tested successfully through a rigorous test program. The ACU will serve as a most frequently used calibration device to meet the challenging physics requirements of JUNO.


\acknowledgments

This work is supported by the Strategic Priority Research Program of
the Chinese Academy of Sciences, Grant No. XDA10010800, and the CAS
Center for Excellence in Particle Physics (CCEPP). We are thankful for
the support from the Office of Science and Technology, Shanghai
Municipal Government (Grant No. 16DZ2260200), and the support from the
Key Laboratory for Particle Physics, Astrophysics and Cosmology,
Ministry of Education.


\end{document}